\documentclass[%
 reprint,
 amsmath,amssymb,
 aps,
]{revtex4-2}

\usepackage{silence}
\usepackage{xcolor}
\usepackage{natbib}
\usepackage{graphicx}% Include figure files
\usepackage{dcolumn}% Align table columns on decimal point
\usepackage{float}
\usepackage{lipsum}
\usepackage[caption=false]{subfig}
\usepackage{amsmath}

\usepackage{bm}% bold math
\usepackage{hyperref}% add hypertext capabilities
\usepackage[mathlines]{lineno}% Enable numbering of text and display math
\begin{document}

\preprint{APS/123-QED}

\title{Measurement of Two-Point Energy Correlators Within Jets in $p$+$p$ Collisions at $\sqrt{s}$ = 200 GeV}

\author{The STAR Collaboration}

\begin{abstract}

Hard-scattered partons ejected from high-energy proton-proton collisions undergo parton shower and hadronization, resulting in collimated collections of particles that are clustered into jets. A substructure observable that highlights the transition between the perturbative and non-perturbative regimes of jet evolution in terms of the angle between two particles is the two-point energy correlator (EEC). In this letter, the first measurement of the EEC at RHIC is presented, using data taken from 200 GeV $p$+$p$ collisions by the STAR experiment.  The EEC is measured both for all the pairs of particles in jets and separately for pairs with like and opposite electric charges.  These measurements demonstrate that the transition between perturbative and non-perturbative effects occurs within an angular region that is consistent with expectations of a universal hadronization regime that scales with jet momentum.  Additionally, a deviation from Monte-Carlo predictions at small angles in the charge-selected sample could result from mechanics of hadronization not fully captured by current models.
\end{abstract}

\maketitle

%\tableofcontents

During a high-energy collision of protons, such as those at the Relativistic Heavy-Ion Collider (RHIC) and Large Hadron Collider (LHC),  hard scatterings between partons within the protons can occur, resulting in large transfers of momentum.  This results in partons leaving the collision with high transverse momenta ($p_{\rm T}$) and virtualities.  In vacuum, this virtuality is shed via splitting \citep{Larkoski:2017bvj}, with the initial splittings able to be described well by perturbative quantum chromodynamics (pQCD).  Later splittings become less perturbative and split particles are separated by increasingly small angles.  This process is referred to as a parton shower and terminates in confinement of the partons into hadrons (hadronization).  The final-state hadrons can be clustered into objects called jets using jet-finding algorithms both experimentally and theoretically \citep{Cacciari:2008gp,Cacciari:2011ma}.

Jet substructure observables derived from jet constituents provide access to information related to the parton shower and hadronization \citep{Chien:2016scy,ATLAS:2018bvp}. Due to the inapplicability of perturbative calculations in the low-energy regime, the exact mechanics of hadronization are less understood than the perturbative part of jet evolution \cite{Wobisch:1998wt} and many observables have been defined in such a way as to limit the impact of non-perturbative effects \citep{Larkoski:2014wba,STAR:2021lvw,STAR:2020ejj}.  In contrast, one way to study hadronization is through substructure observables that can isolate non-perturbative and perturbative regimes as well as the transition between them to separate angular regions.  As the parton shower is angular-ordered and well described by DGLAP evolution equations \citep{Andersson:1983ia,Altarelli:1977zs,Hoche:2017hno}, the angular scale at which a perturbative framework ceases to fully describe the parton shower is identifiable.  One such observable is the $N$-point energy correlator, which describes energy flow among $N$ particles as a function of an angular scale, $R_{L}$.  This scale is defined as the largest distance in angular space between any two of the $N$ particles involved in the correlation, $R_{L} = \sqrt{\Delta\eta^2+\Delta\phi^2}$, where $\eta$ is the pseudorapidity and $\phi$ is the azimuthal angle.  The simplest case, the two-point energy correlator (EEC) describes the energy flow within a jet in terms of 1$\rightarrow$2 splittings.  The EEC distribution is expected to be separated into three regimes: i) the free diffusion of hadrons at small angles, ii) the perturbative evolution of the parton shower at large angles, and iii) the transition region between them.  This results in the EEC distribution scaling linearly with $R_{L}$ at small angles and $\propto 1/{R_{L}}^{1-\gamma_2}$ at large angles, where $\gamma_2$ is the anomalous dimension governing the EEC, in this case the third moment of the splitting function \citep{Chen:2020vvp}.  The transition regime, where neither non-perturbative nor perturbative effects completely dominate, occurs around the peak of the EEC distribution. 

The transition in the EEC is also expected to occur at smaller angles $\propto \Lambda_{\rm QCD}/p_{\rm T,jet}$ \citep{Chen:2020vvp,Komiske:2022enw} for jets with higher transverse momenta ($p_{\rm T, jet}$), for a given initiator flavor, $i.e.$ a quark or gluon that initiates the jet \cite{Craft:2022kdo}. As highly virtual gluons typically fragment wider than quarks, this transition region will emerge at larger angles for a gluon-initiated jet than for a quark-initiated jet.  Due to the lower center-of-mass energies at RHIC relative to the LHC, EEC measurement at RHIC energies is able to capture low-momentum jets, which have a larger angular phase space dominated by non-perturbative effects, while maintaining a high fraction of (light) quark-initiated jets \citep{deFlorian:2007fv}.  This quark jet fraction increases with increasing $p_{\rm T, jet}$ at RHIC, which may slightly affect the location of the transition region.   

Details of the hadronization process are further probed via measuring the EEC in terms of hadron pairs that have like or opposite electric charge to inform how charge correlations are carried through the hadronization process \citep{Lee:2023npz}.  The data can then be compared with Monte-Carlo generators using differing models of hadronization, such as string breaking in PYTHIA 8 \citep{Adkins:Thesis} and cluster hadronization in HERWIG 7 \citep{Marchesini:1991ch}. String breaking hadronization models are expected to enhance correlations of opposite-sign pairs at small angles, as a broken string forms an opposite-charge pair at a small separation. In this letter, we will first outline the experimental methods and definitions used to perform jet substructure measurements.  Next, we will describe the correction procedure designed to remove detector effects from the measurement.  Finally, we report the first experimental measurement of the EEC at RHIC in proton-proton ($p$+$p$) collisions for several jet momentum and pair charge selections.

The data used in this analysis were collected by STAR \citep{STAR:2002eio} in 2012, from $p$+$p$ collisions at the center-of-mass energy $\sqrt{s} = $ 200 GeV using an online trigger designed to create a jet-enriched sample, which resulted in an integrated luminosity of 22.9 pb$^{-1}$.  This trigger requires an energy deposit of 7.3 GeV in at least one of 18 overlapping Barrel Electromagnetic Calorimeter (BEMC) patches spanning $1 \times 1$ in $\eta - \phi$ space. The BEMC, consisting of 4800 towers each having a size of $0.05 \times 0.05$ in $\eta - \phi$ space, can measure electromagnetic energy deposits, such as those from photons and electrons \citep{STAR:2002ymp}. Additionally, STAR can reconstruct both angular and momentum information of charged particles with its Time Projection Chamber (TPC) \citep{Anderson:2003ur}. 
  When calculating charged particle energy, the charged pion mass is assumed \citep{ALICE:2017nij}.  The BEMC tower energy is set equal to the energy deposited in the tower and subtracted by 100$\%$ of the transverse momenta of all TPC tracks projected to hit that tower. Towers with negative resulting energies are discarded.  This subtraction process is referred to as hadronic correction. Energy deposits in the BEMC are treated as contributions from neutral particles with zero mass for the purpose of jet finding.  Both the TPC and BEMC allow for a pseudorapidity acceptance of $|\eta| < 1$ and full azimuthal coverage. For this analysis, we require both tracks and towers with $0.2 < p_{\rm T} < 30$ GeV/$c$. Other event-level and track-level selections are the same as previous STAR jet substructure analyses \citep{STAR:2020ejj,STAR:2021lvw}. 
  
  Jets are reconstructed using the anti-$k_{\rm T}$ algorithm \cite{Cacciari:2008gp} with a radius of $R_{\rm jet} =$ 0.6 and with both charged tracks and neutral energy deposits as inputs.  Additionally, results for jets reconstructed using a radius of $R_{\rm jet} =$ 0.4 are shown in the Appendix. The angular resolution provided by tracking in the TPC is superior to that provided by energy deposits in the BEMC.  Consequently, while jets are reconstructed using both charged tracks and BEMC deposits, only charged tracks are examined for determination of the EEC. Found jets are restricted to a pseudorapidity space of $|\eta_{\rm jet}| \leq  1-R_{\rm jet}$ to avoid partially reconstructed jets. This analysis is only presented for jets with $p_{\rm T, jet} > 15$ GeV/$c$ in order to minimize the bias towards jets with an enhanced neutral energy fraction resulting from the trigger.  Additionally, no attempt is made to remove the underlying event from the analysis, as it is expected to be negligible for the selected jet sample. 

Experimentally, the EEC is defined in Eq. \ref{EEC_equation}. The energy weighting of each pair is defined as the energy product of the involved tracks divided by the square of jet transverse momentum, $E_{i}E_{j}/p_{\rm T,jet}^2$, with $i$ and $j$ representing two distinct tracks within a jet and $R_{L}$ being their angular separation.

\begin{equation}
\text{Normalized EEC} = \frac{1}{\sum\limits_{\mathrm jets}\sum\limits_{i \neq j}\left(\frac{{E_i} {E_j}}{p_{\rm T,jet}^2}\right)} \frac{\mathrm{d}\left(\sum\limits_{\mathrm jets}\sum\limits_{i \neq j}\left(\frac{{E_i} {E_j}}{p_{\rm T,jet}^2}\right)\right)}{\mathrm{d} R_{L}}
\label{EEC_equation}
\end{equation} 

  \noindent.  The integral of this distribution is normalized to unity in order to highlight shape changes across kinematic selections. 

\begin{figure}[ht!]
\centering
\begin{center}
\includegraphics[width=1\linewidth]{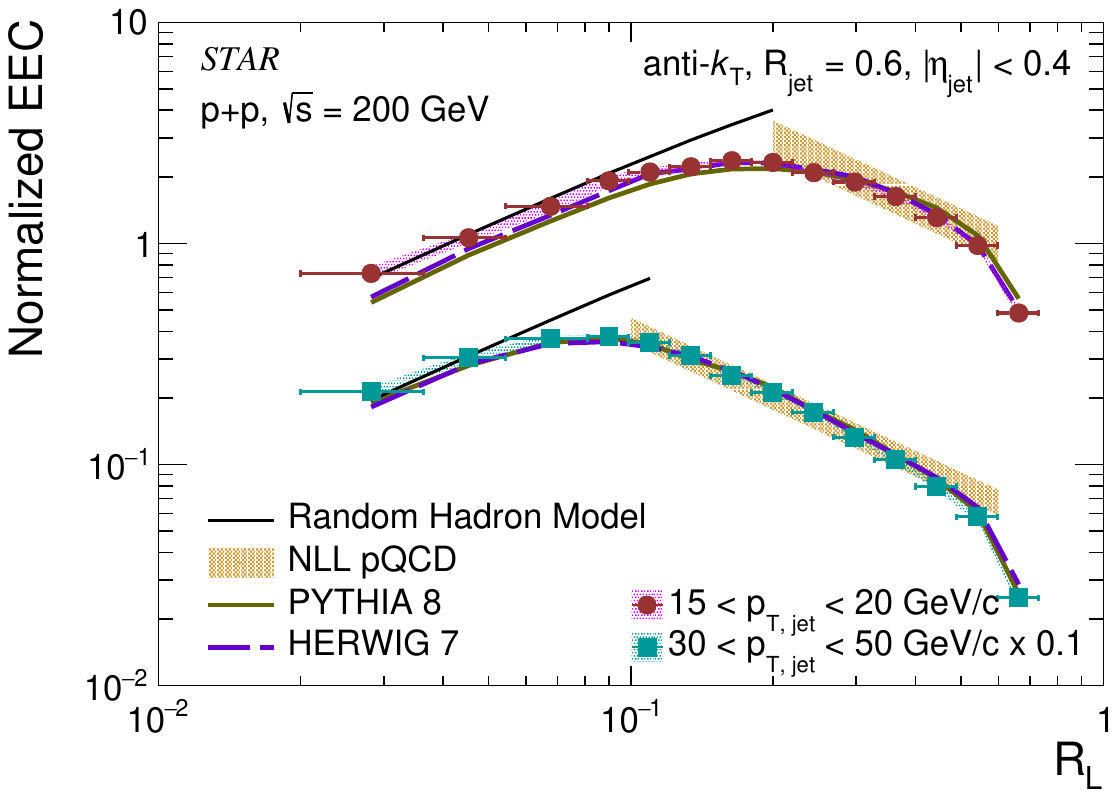}
\end{center}
\caption{Corrected distributions of the normalized EEC differential in $R_{L}$ for $R_{\rm jet}=$ 0.6, with jet transverse momentum selections 15 $< p_{\rm T, jet} <$ 20 GeV/$c$ and 30 $< p_{\rm T, jet} <$ 50 GeV/$c$ (scaled for comparison). Random hadron scaling at small angles and next-to-leading-logarithmic-pQCD (NLL pQCD) \citep{Lee:2022uwt} calculations at large angles are presented alongside Monte-Carlo predictions \citep{Marchesini:1991ch,Aguilar:2021sfa}.}
\label{Tamis_Figure1}
\end{figure}

Correction for detector effects uses $p$+$p$ events generated at $\sqrt{s} =$ 200 GeV via the PYTHIA 6 STAR-Tuned event generator, referred to as the particle-level sample \citep{Adkins:Thesis}. These events are then used to create a detector-level sample by propagating particles through a GEANT-3 \citep{Brun:1082634} simulation of the STAR detector and embedding them within events that are collected by triggering randomly during beam crossings. These events are then analyzed identically to data, including the enforcement of the requirement on BEMC patch energy derived from simulated tower energy deposits. Comparisons of this detector-level simulation with uncorrected data show reasonable agreement, which provides justification for using PYTHIA to inform corrections. Particle-level and detector-level jets are considered to be matched if their axes are within one jet radius of each other, $i.e.$ $\sqrt{(\Delta\eta_{jet})^2+(\Delta\phi_{jet})^2} \leq R_{\rm jet}$, the jet pair is determined to be matched.  If multiple candidates for matching are available at the detector level, the one closer in momentum to the particle-level jet is taken.  Tracks within the jets are then matched on a similar basis, with $\sqrt{(\Delta\eta_{\rm track})^2+(\Delta\phi_{\rm track})^2} \leq 0.01$. Both jets and tracks can be missed in the detector-level sample for various reasons, including inefficiencies in trigger selection and track reconstruction.  If either track involved in a correlation is not reconstructed in the detector-level sample, it is determined to be missed.  Likewise, correlations that are only reconstructed at detector level are determined to be fake.  Fully matched, missed, and fake correlations then inform correction via an iterative Bayesian unfolding procedure \citep{DAgostini:2010hil}.  The RooUnfold software package \citep{Adye:2011gm} is used to create a multi-dimensional response matrix mapping particle-level jets to detector-level jets. Due to the excellent $R_{L}$ resolution, its correction is expected to be negligible, leaving the largest correction as that due to $p_{\rm T, jet}$ smearing, which biases jets towards lower momenta.  The energy weighting is also directly dependent on the jet momentum and behaves differently for different $R_{L}$.  For this reason, a Bayesian unfolding \citep{Wobisch:1998wt} is performed in three dimensions: $p_{\rm T, jet}$, $R_{L}$, and the energy weight $E_{i} E_{j}/p_{\rm T, jet}^{2}$.  Bayesian unfolding iteratively infers the particle-level distribution that would generate the measured detector-level data based on the response matrix.  This procedure is then carried out to a nominal 4 iterations and validated by a split-sample closure test, where the response matrix is built using one statistically independent half of the embedding and used to unfold the other half.
Several sources of systematic uncertainty are estimated by varying parameters used in the generation of the detector-level sample used for unfolding. This includes varying the BEMC tower energy by its uncertainty of 3.8$\%$ as well as the tracking efficiency of the TPC by its uncertainty of 4$\%$. The hadronic correction is varied between 100$\%$ and 50$\%$ of matched track $p_{\mathrm T}$. The uncertainty associated with the unfolding process itself is estimated by varying the number of unfolding iterations and the prior, the later of which is achieved by re-weighting the PYTHIA+GEANT3 embedding sample according to HERWIG \citep{STAR:2021lvw,STAR:2020ejj}.

\begin{figure} [ht!]
\centering
    \begin{center}  
    
\includegraphics[width=1.0\linewidth]{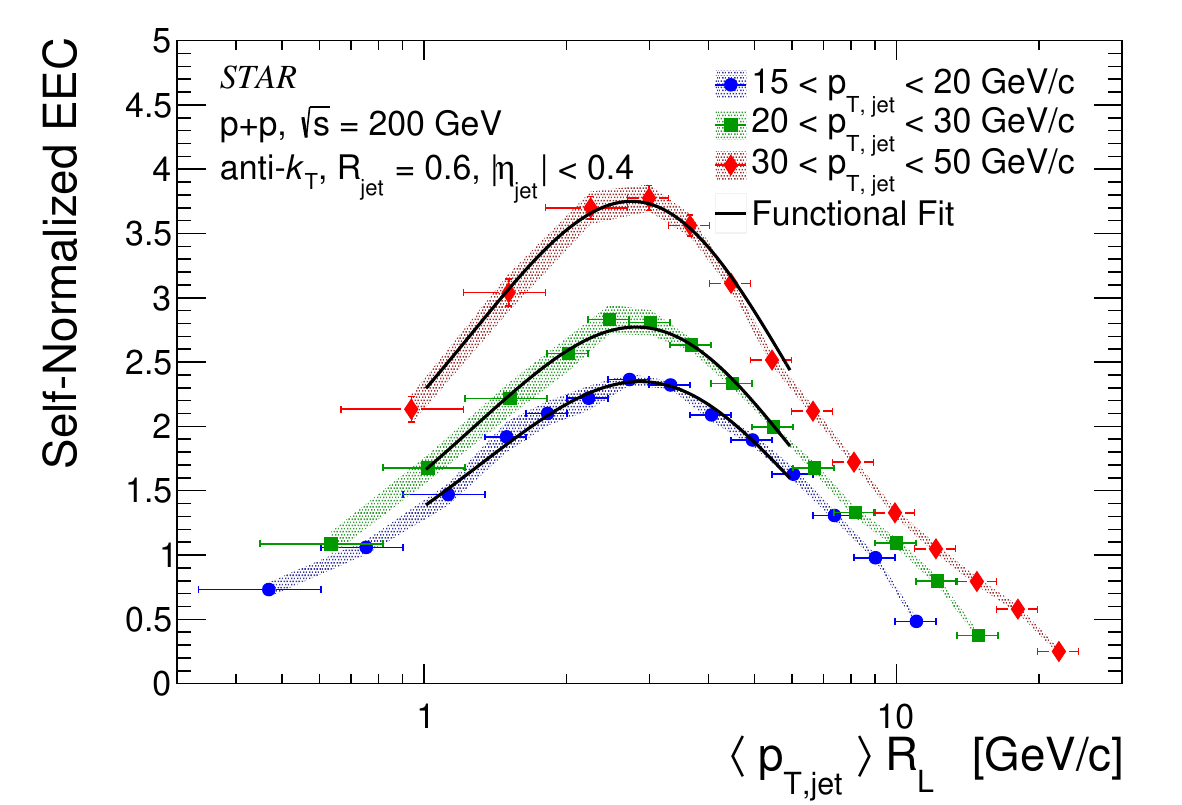}
\caption{Corrected distributions of the normalized EEC within jets, differential in $ \left\langle p_{\rm T,jet} \right\rangle R_{L} $ at $R_{\rm jet} =$ 0.6 for several $p_{\rm T, jet}$ selections. Each distribution is normalized to integrate to one in $R_{L}$ and fitted with the function described in Eq. \ref{FitEquation} prior to shifting.  The resulting fit is also shifted and compared with data within the range $1 < \left\langle p_{\rm T,jet} \right\rangle R_{L} < 6$ GeV/$c$.}
\label{Tamis_Figure2}
\end{center}
\end{figure}

The EEC for two ranges of jet transverse momentum, $p_{\rm T, jet}$ $\in$ [15, 20] and [30, 50] GeV/$c$ (scaled by 0.1 for visual clarity), using $R_{\rm jet} = 0.6$ are shown in Fig. \ref{Tamis_Figure1}, with systematic uncertainties represented by checkered bands, theoretical predictions represented by striped bands, and model predictions by solid curves. Moving from smaller to larger angles, the behavior of each distribution is separated into a non-perturbative free hadron region, a transition region, and a perturbative region, as predicted in Ref. \citep{Komiske:2022enw}. Particle pairs at angles smaller than the transition region likely correspond to two particles that are correlated post hadronization, where energy freely diffuses in a hadron gas. Therefore, the hadronic region is observed to increase linearly as a function of $R_{L}$, corresponding to a uniform distribution of energy. The scaling within this region compares well to a toy model that distributes hadrons, generated from an exponentially falling $p_{\rm T}$ spectrum, uniformly in angle on a sphere.  This represents the simplistic case of energy being uniformly distributed with no correlation.  At large angles, next-to-leading-logarithmic-pQCD calculations (NLL pQCD) \citep{Lee:2022uwt} can directly be compared to data and are shown alongside the data in Fig. \ref{Tamis_Figure1}.  These predictions are calculated for ranges between (3 GeV/$c$)/$p_{\rm T, low} < R_{L} < R_{\rm jet} $, where $p_{\rm T, low}$ is the lower bound of a given $p_{\rm T, jet}$ selection.  Correlations farther than the jet radius fall off due to geometric effects, as many correlations at this distance are cut out of the sample due to the imposition of a jet radius. As the distribution agrees well both with the theoretical predictions of the perturbative regime and the random hadron model in the non-perturbative region, the start and end of the transition regime can be defined as where the distribution breaks from these scalings on either side.  This is motivated by the expectation that the transition region encompasses the shift from the perturbative to the non-perturbative regime in the parton shower. As $p_{\rm T, jet}$ increases from [15, 20] to [30, 50] GeV/$c$, this transition region moves to smaller angles, $i.e.$ later in jet evolution. Additionally, comparisons with both the PYTHIA 8 Detroit Tune \cite{Aguilar:2021sfa} and the HERWIG 7.2 default tune \citep{Marchesini:1991ch} are presented, and they both describe the data well within systematics in the [30, 50] GeV/$c$ bin. However, PYTHIA predicts a transition region at a larger angle than is seen in data for the [15, 20] GeV/$c$ bin, resulting in a slightly under-predicted hadronic region, while HERWIG shows deviation in the leftmost two bins. 

The movement of the transition region as a function of jet momentum can be examined more closely in Fig. \ref{Tamis_Figure2},  where the EEC distribution is scaled by average $p_{ \rm T, jet}$ to create a distribution differential in $\left\langle p_{\rm T,jet} \right\rangle R_{L}$. $\left\langle p_{\rm T,jet} \right\rangle$ is the average jet momentum in a given momentum range, as determined via PYTHIA 6: 16.7, 22.7, and 33.5 GeV/$c$ for the [15, 20], [20, 30], and [30, 50] GeV/$c$ intervals, respectively.  Scaled as such, the peak position, characterizing the transition region, exhibits a universal scale independent of jet momentum.  In order to characterize this scale, a fit to the transition region using Eq. \ref{FitEquation} \cite{IanPrivateCommunication} is performed:  

\begin{equation}
\begin{aligned}
F\left( R_{L}\right) =  C \frac{R_{L}}{{\left( R_{L}^2+ T \right)}^{3/2}}
\label{FitEquation}
\end{aligned}
\end{equation}

\noindent, where $C$ and $T$ are free parameters.  This function describes a smooth crossover between a linear scaling at $R_{L} \ll T$ and $\sim{R_{L}}^{-2}$ scaling at $R_{L} \gg T$. The location of the transition region is determined only by the dimensionless parameter $T$.  Therefore, this parameter can be used to quantify the change in the transition region as a function of jet momentum. Writing the function in terms of $\left\langle p_{\rm T,jet} \right\rangle R_{L}$, one can expect the dimensionful parameter ${\left\langle p_{\rm T,jet} \right\rangle}^2 T$ to be constant regardless of jet momentum.  The resulting fits are shown in Fig. \ref{Tamis_Figure2} with corresponding ${\left\langle p_{\rm T,jet} \right\rangle}^2 T$ values of $16.3\pm 0.7$, $15.6\pm 0.8$, and $15.0\pm 1.0$ (GeV/$c$)\textsuperscript{2} for $p_{\rm T, jet}$ $\in$ [15, 20], [20, 30], and [30, 50] GeV/$c$, respectively. The peak of the distribution is located at $\left\langle p_{\rm T,jet} \right\rangle R_{L}$ = $\sqrt{\left\langle p_{\rm T,jet} \right\rangle^2 T/2}$.  The invariance with $p_{\rm T, jet}$ is inline with the theoretical expectation of a universal energy scale.

\begin{figure}[ht!]
  \centering
  \includegraphics[width=8cm]{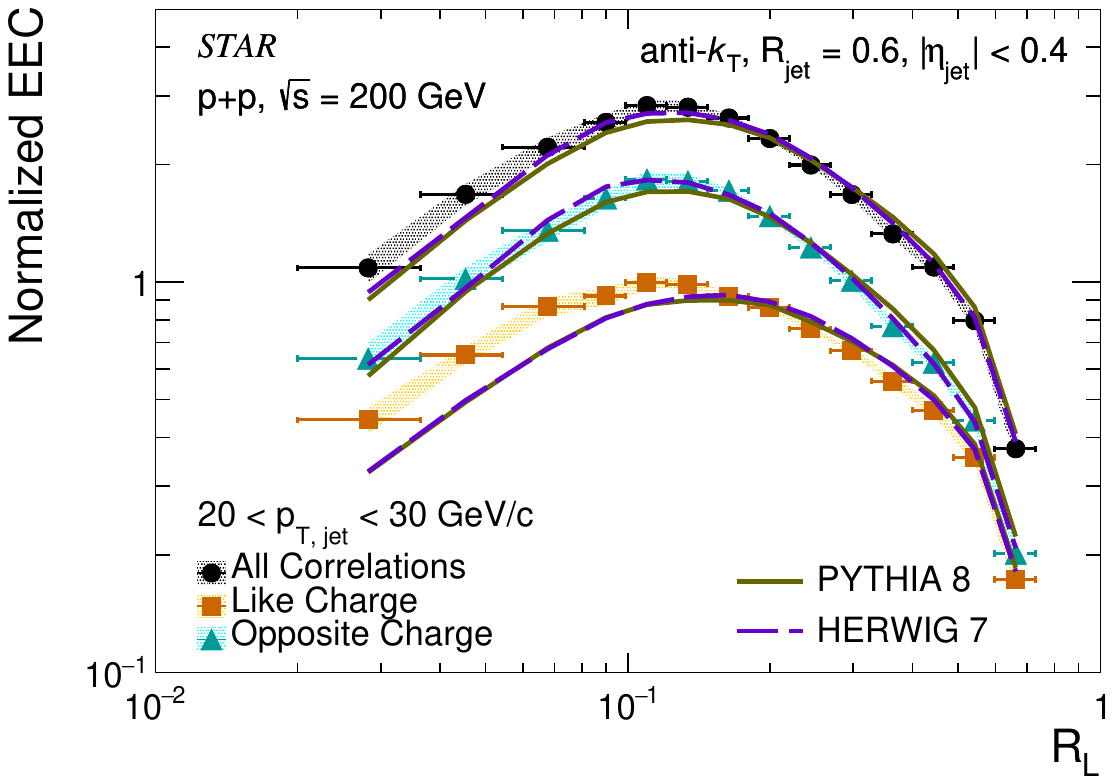}
  \includegraphics[width=8cm]{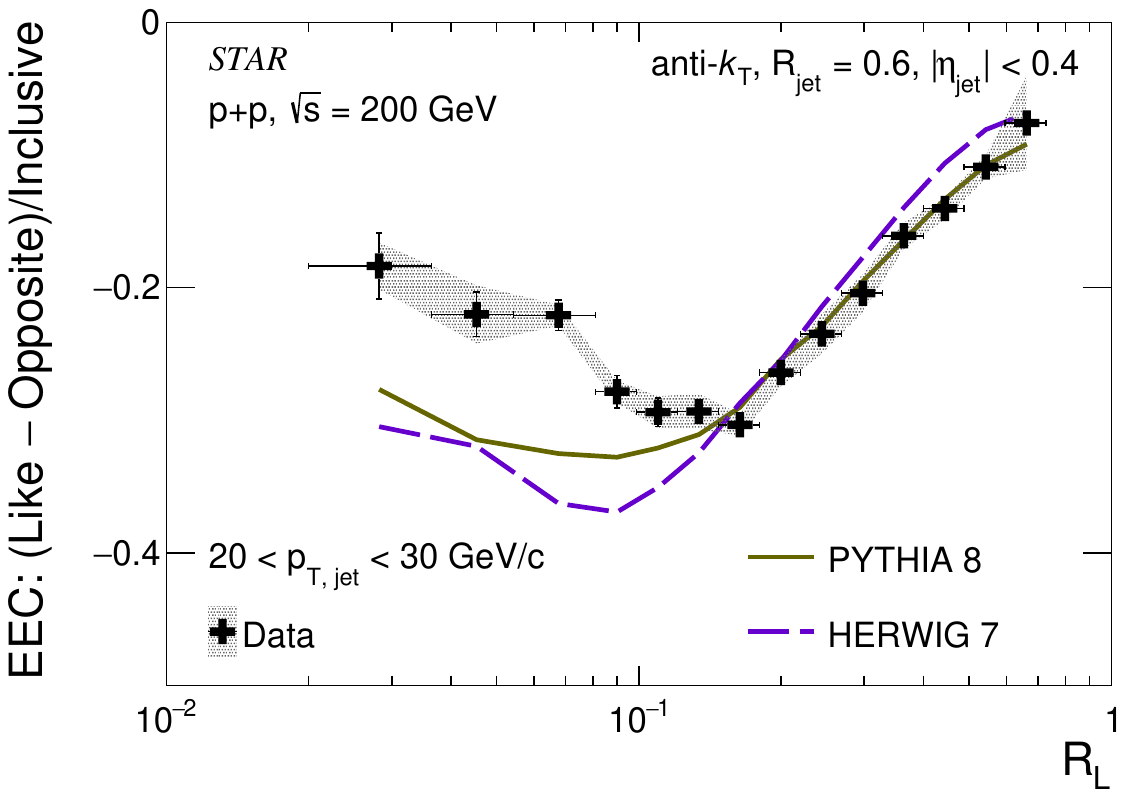}
\caption{Corrected distributions of the charge-selected EEC compared with the inclusive case (top panel) and charged ratio (bottom panel) for $R_{\rm jet}$ = 0.6 with a jet transverse momentum selection of 20  $ < p_{\rm T, jet} <$ 30 GeV/$c$.  Comparisons with the PYTHIA 8 Detroit Tune \citep{Aguilar:2021sfa} and the default tune of HERWIG 7 are given \citep{Marchesini:1991ch}.}
\label{Tamis_Figure3}
\end{figure}

As EECs are sensitive to hadronization, a possible extension is to examine the correlations between pairs of charged hadrons with selections on their relative charges. Although the inclusive EEC shows results in the hadronization regime consistent with randomly diffusing hadrons, correlations can be introduced in the charge of these hadrons which can probe how charge conservation is enforced at hadronization. The ratio of cross-sections of like- and opposite-charge pairs is demonstrated to depend on the hadronization model \citep{Chien:2021yol}. While the general EEC distribution is observed to be captured well by PYTHIA and HERWIG, it is important to test an observable with additional sensitivity to hadronization to further elucidate the validity of hadronization models used by Monte-Carlo generators. This is examined by constructing the EEC as previous, but including the charge product in the energy weighting \citep{Lee:2023npz}.  This is experimentally defined in Eq. \ref{chargedEEC_Equation}: 

\begin{equation}
\begin{aligned}
\text{Charge-weighted EEC}  &= \frac{\mathrm{d}\left(\sum\limits_{\mathrm{jets}}\sum\limits_{i \neq j}\frac{{E_i Q_i} {E_j Q_j}}{p_{\rm T,jet}^2}\right)}{\mathrm{d} R_{L}} \\&= \mathrm{EEC}_{\rm like}-\mathrm{EEC}_{\rm opposite}
\label{chargedEEC_Equation}
\end{aligned}
\end{equation}

.  In the case of the two-point correlator, this separates the distribution into two components: one constructed using only like-charge pairs and one using only opposite-charge pairs.  For this analysis, the like-charge and opposite-charge EECs are corrected separately and normalized relative to the inclusive distribution.  Both distributions are displayed alongside the inclusive EEC in the top panel of Fig. \ref{Tamis_Figure3}.  Both PYTHIA and HERWIG under-predict the relative contribution of like-charge correlations for angles smaller than the transition region for $20 < p_{\rm T, jet} < 30$ GeV/$c$.  Additionally, both models predict that the peak of the distribution will shift to larger angles for like-charge correlations and smaller angles for opposite-charge correlations, a trend that is not observed in data, where both subsamples have a peak consistent with the inclusive sample.

In order to provide a more quantitative description of the charge correlations, subtracting these two distributions forms the charge-weighted distribution in Eq. \ref{chargedEEC_Equation}.  This can then be divided by the inclusive EEC to form a ratio describing the magnitude of charge correlation on a scale from $-$1 to 1.  A value of 0 would be expected for an infinite thermal bath with equal probability to form a like-charge or opposite-charge correlation.  In this scheme, changes in the shape of the ratio can highlight increased correlations with respect to certain angular distances, with a more negative value corresponding to increased correlation of opposite-charge pairs.  This ratio is shown in the bottom panel of Fig. \ref{Tamis_Figure3}.  While opposite-charge correlations are overall more prevalent than like-charge correlations due to charge conservation, their relative abundance varies across the $R_{L}$ range.  At angles larger than the transition region, the slope is positive against $R_{L}$, showing that charge becomes de-correlated at larger angles.  This can be caused by more opposite-sign correlations being formed as the shower progresses. This trend is correctly described by both Monte-Carlo models, although it is captured more accurately by PYTHIA.  At angles below the transition regime, the slope begins to reverse, and the ratio moves closer to zero, which reflects a charge de-correlation that is not predicted by Monte-Carlo models. This points to a charge-dependent aspect of hadronization that is not fully captured by current Monte-Carlo models, providing opportunity for further model development.

%Tables~\ref{tab:table1}, \ref{tab:table3}, \ref{tab:table4}, and \ref{tab:table2}%
%\begin{table}[b]
%\caption{\label{tab:table2}
%A table with numerous columns that still fits into a single column. 
%Here, several entries share the same footnote. 
%Inspect the \LaTeX\ input for this table to see exactly how it is done.}
%\begin{ruledtabular}
%\begin{tabular}{cccccccc}
% &Value \\
%\hline
%CMS & 1\\
%ALICE & 1 \\
%STAR & 1
%\end{tabular}
%\end{ruledtabular}
%\footnotetext[1]{Here's the first, from Ref.~\onlinecitep{feyn54}.}
%\footnotetext[2]{Here's the second.}
%\footnotetext[3]{Here's the third.}
%\footnotetext[4]{Here's the fourth.}
%\footnotetext[5]{And etc.}
%\end{table}

In this letter, the first measurement of the two-point energy correlator (EEC) at RHIC has been presented using STAR data from $p$+$p$ collisions at $\sqrt{s}$ = 200 GeV, showing clear separation of the jet evolution into three regimes, a free hadron regime, a perturbative regime associated with quarks and gluons, and the transition between the two. Agreement is observed with theoretical predictions in the region dominated by perturbative effects as well as with scaling expectations of the free hadron regime, allowing for identification of the transition regime. 
 When scaled by jet momentum, the peak of the transition region can be characterized by a scale ${\left\langle p_{\rm T,jet} \right\rangle}^2 T \simeq$ 15.5 (GeV/$c$)\textsuperscript{2} independent of jet momentum. This is in line with the expectation: a universal hadronization scale that causes higher energy jets to hadronize at later times.  Comparisons with other experiments, such as the studies performed by CMS and ALICE, will allow for studying the effects of changing initiator flavor due to different collision energies.  Additional studies across experiments with identified initiating parton flavor, such as photon-jet measurements or heavy flavor tagging, can be used to inform the commonality of this transition against the collision energy.  Furthermore, separating the track pairs with like-charge and opposite-charge correlations allows for increased sensitivity to hadronization.  At angles below the transition regime, predictions from both PYTHIA and HERWIG fail to fully describe the charge-weighted EEC. This provides an opportunity for further development of Monte-Carlo generators in modeling the hadronization process.  Since the EEC allows for time-scale separation in jets, it could prove useful in probing the modifications of jet evolution in the quark gluon plasma created in heavy-ion collisions \citep{Andres:2023xwr,Barata:2023bhh,Bossi:2024qho,Rai:2024ssx}. This study, therefore, could serve as a baseline in future measurements of the EEC in heavy-ion collisions at RHIC.  
 
We thank the RHIC Operations Group and SDCC at BNL, the NERSC Center at LBNL, and the Open Science Grid consortium for providing resources and support.  This work was supported in part by the Office of Nuclear Physics within the U.S. DOE Office of Science, the U.S. National Science Foundation, National Natural Science Foundation of China, Chinese Academy of Science, the Ministry of Science and Technology of China and the Chinese Ministry of Education, the Higher Education Sprout Project by Ministry of Education at NCKU, the National Research Foundation of Korea, Czech Science Foundation and Ministry of Education, Youth and Sports of the Czech Republic, Hungarian National Research, Development and Innovation Office, New National Excellency Programme of the Hungarian Ministry of Human Capacities, Department of Atomic Energy and Department of Science and Technology of the Government of India, the National Science Centre and WUT ID-UB of Poland, the Ministry of Science, Education and Sports of the Republic of Croatia, German Bundesministerium f\"ur Bildung, Wissenschaft, Forschung and Technologie (BMBF), Helmholtz Association, Ministry of Education, Culture, Sports, Science, and Technology (MEXT), Japan Society for the Promotion of Science (JSPS) and Agencia Nacional de Investigaci\'on y Desarrollo (ANID) of Chile.


%bibliographystyle{unsrt}
\bibliography{apssamp}% Produces the bibliography via BibTeX.

\clearpage
\section{Appendix}

\begin{figure}[ht!]
\centering
\begin{center}
\includegraphics[width=0.9\linewidth]{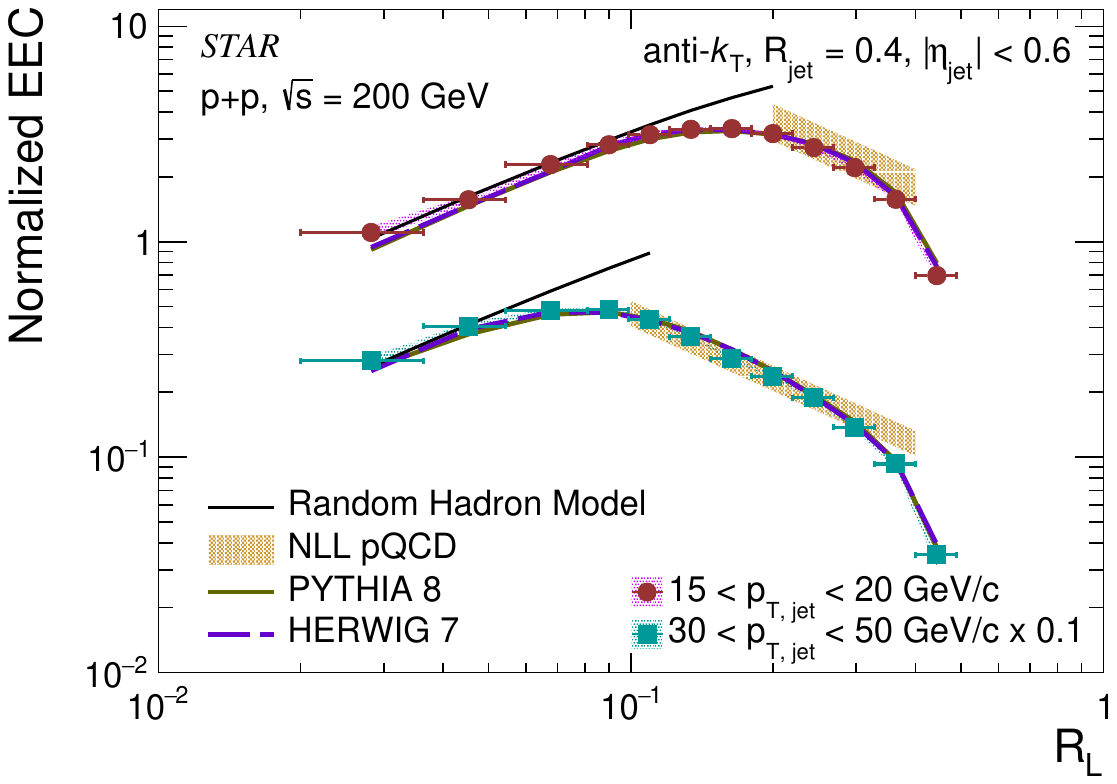}
\end{center}
\caption{Corrected distributions of the normalized EEC differential in $R_{L}$ for $R_{\rm jet}=$ 0.4, with jet transverse momentum selections 15 $< p_{\rm T, jet} <$ 20 GeV/$c$ and 30 $< p_{\rm T, jet} <$ 50 GeV/$c$ (scaled for comparison). Random hadron scaling at small angles and next-to-leading-logarithmic-pQCD (NLL pQCD) \citep{Lee:2022uwt} calculations at large angles are presented alongside Monte-Carlo predictions \citep{Marchesini:1991ch,Aguilar:2021sfa}.}
\label{Tamis_Figure1_0p4}
\end{figure}

While jets are reconstructed using only a radius of $R_{\rm jet} =$ 0.6 in the main body of the paper, the analysis is repeated for jets reconstructed with $R_{\rm jet}$ = 0.4 using the same analysis procedure, as in the main text.  In vacuum, performing jet-finding with larger radii allows wider-fragmenting gluon jets to recover more of their energies and pass $p_{\rm T}$ selections more easily.  This increases the fraction of gluon-initiated jets in the resulting sample, which could potentially cause the transition region to emerge at smaller angles for $R_{\rm jet}$ = 0.4 jets relative to $R_{\rm jet}$ = 0.6 jets.  The EEC is shown for $R_{\rm jet}$ = 0.4 for two selections of jet transverse momentum in Fig. \ref{Tamis_Figure1_0p4}.  The distribution is separated into a non-perturbative free hadron region, a transition region, and a perturbative region, similar to that shown in Fig. \ref{Tamis_Figure1}.  The transition region is also observed to move to a smaller angle as jet momentum increases.

\begin{figure}[ht!]
\centering
    \begin{center}  
    
\includegraphics[width=1.0\linewidth]{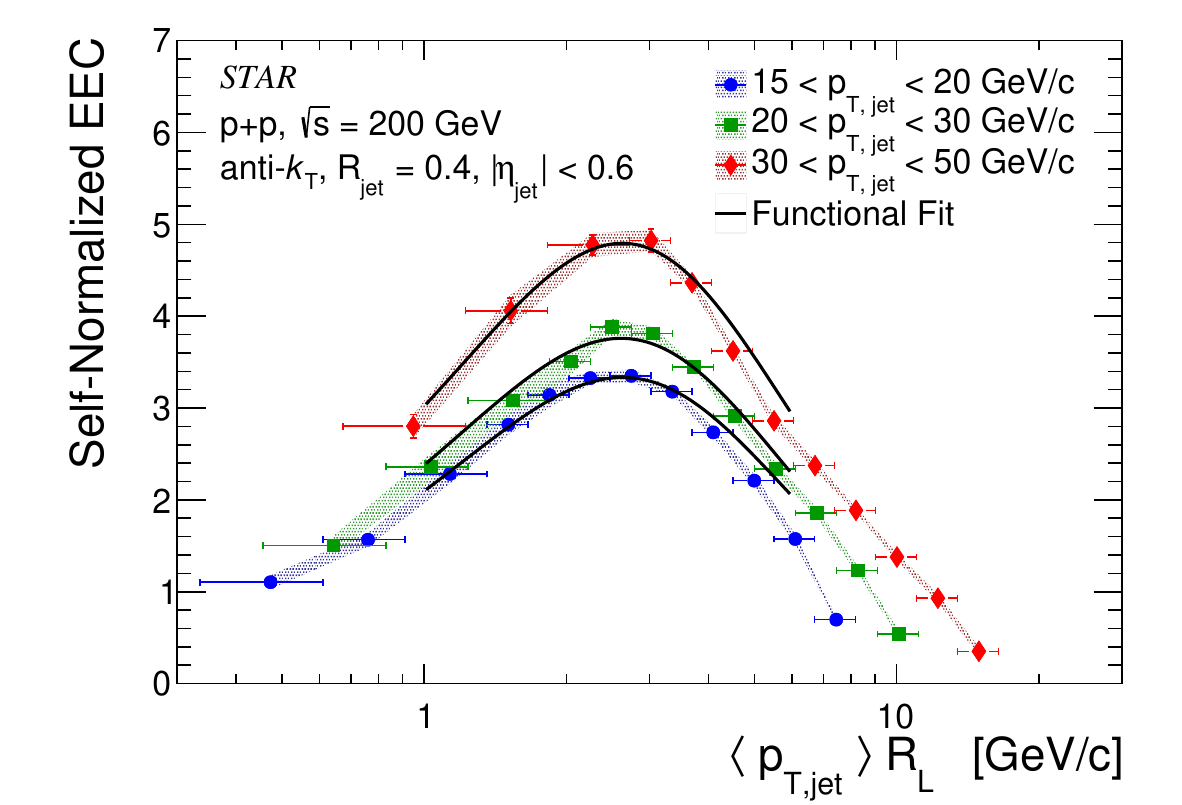}
\caption{Corrected distributions of the normalized EEC within jets, differential in $ \left\langle p_{\rm T,jet} \right\rangle R_{L} $ at $R_{\rm jet} =$ 0.4 for several $p_{\rm T, jet}$ selections. Each distribution is normalized to integrate to one in $R_{L}$ and fitted with the function described in Eq. \ref{FitEquation} prior to shifting.  The resulting fit is also shifted and compared with data within the range $1 < \left\langle p_{\rm T,jet} \right\rangle R_{L} < 6$ GeV/$c$.}
\label{Tamis_Figure2_0p4}
\end{center}
\end{figure}

As is demonstrated in Fig. \ref{Tamis_Figure2}, presenting the EEC differential in $\left\langle p_{\rm T,jet} \right\rangle R_{L}$ collapses the transition region to a consistent scale.  This is shown for the $R_{\rm jet}$ = 0.4 case in Fig. \ref{Tamis_Figure2_0p4}. Determined $\left\langle p_{\rm T,jet} \right\rangle$ values are 16.8, 22.7, and 33.6 GeV/$c$, for $p_{\rm T, jet}$ $\in$ [15, 20], [20, 30], and [30, 50] GeV/$c$, respectively.

To characterize the transition region, the data are fit using Eq. \ref{FitEquation}, with resulting ${\left\langle p_{\rm T,jet} \right\rangle}^2 T$ values of $13.8\pm 0.7$, $13.6\pm 0.6$, and $13.7\pm 1.1$ (GeV/$c$)\textsuperscript{2} for $p_{\rm T, jet}$ $\in$ [15, 20], [20, 30], and [30, 50] GeV/$c$, respectively.  These values are consistent with each other within errors, as well as within 1-2 $\sigma$ of those determined for the $R_{\rm jet}$ = 0.6 case.  Averaging the three values from each radius selection results in a mean value of $13.72 \pm 0.48$ (GeV/$c$)\textsuperscript{2} for $R_{\rm jet}$ = 0.4 and $15.63 \pm 0.48$ (GeV/$c$)\textsuperscript{2} for $R_{\rm jet}$ = 0.6, with a separation of 2.8 $\sigma$ between the two average values.  This is inline with an increased fraction of gluon-initiated jets within a given $p_{\rm T, jet}$ range for jets of larger radius, which would result in a larger value of ${\left\langle p_{\rm T,jet} \right\rangle}^2 T$ for $R_{\rm jet}$ = 0.6 jets \citep{Craft:2022kdo}.

\begin{figure}[ht!]
  \centering
  \includegraphics[width=8cm]{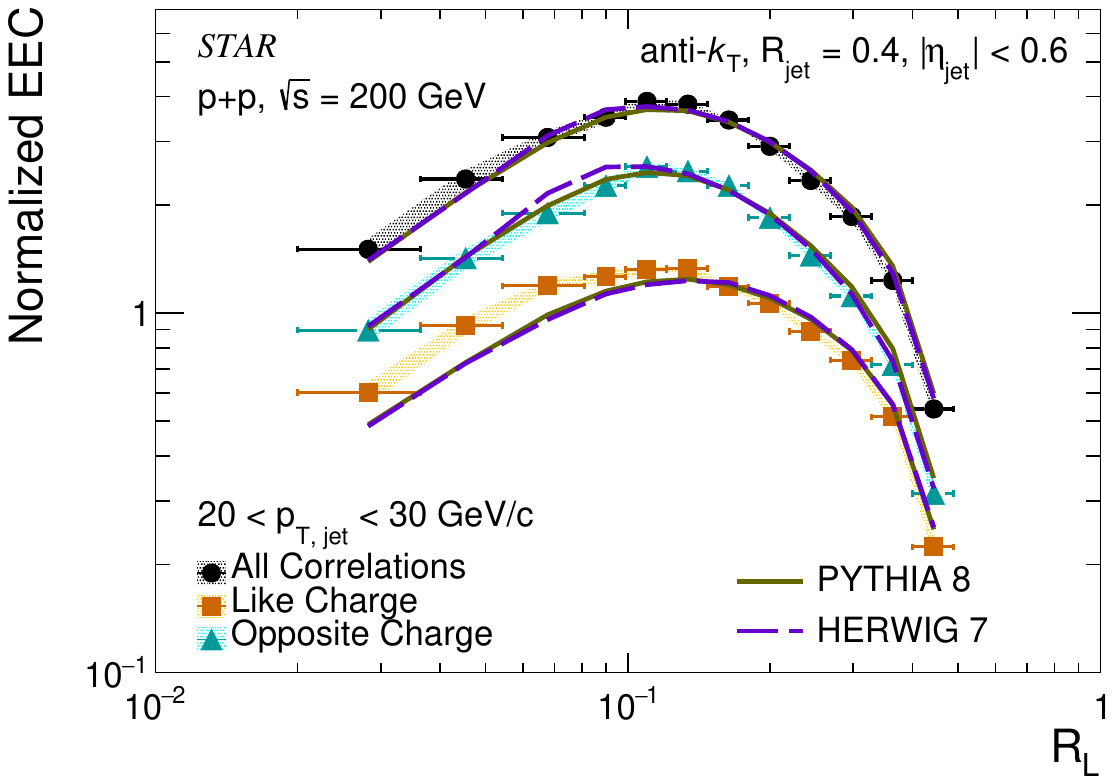}
  \includegraphics[width=8cm]{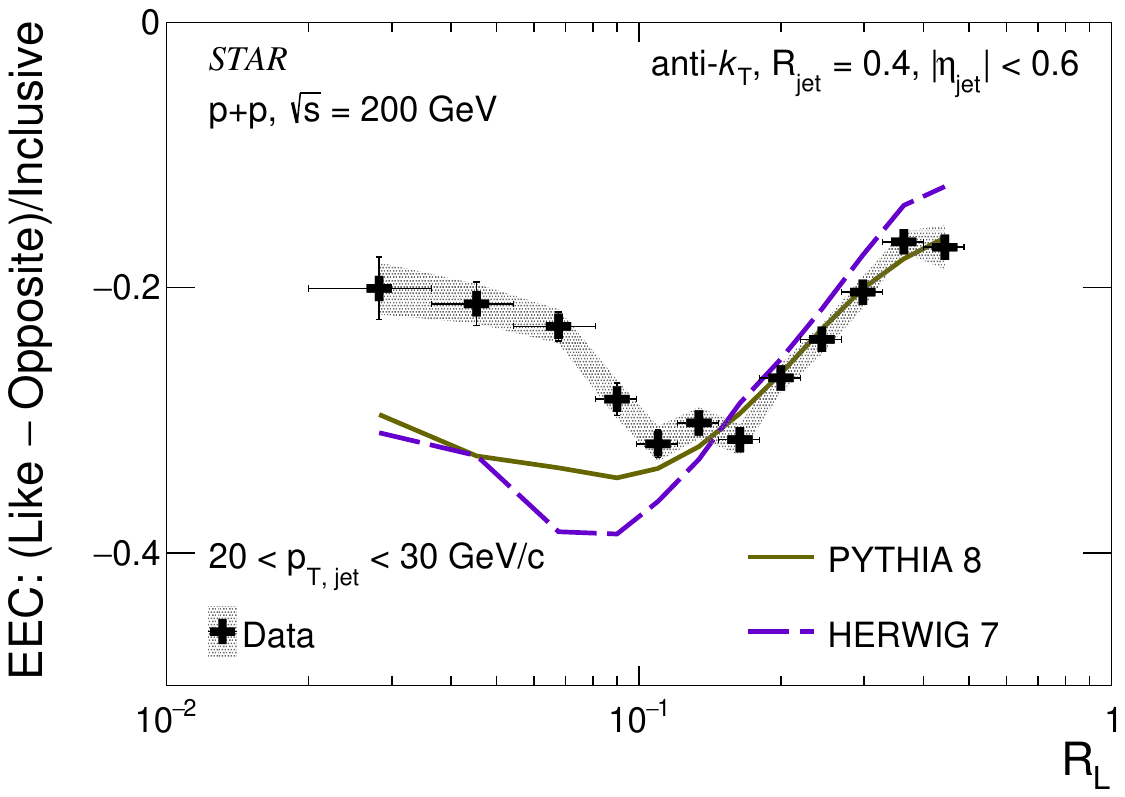}
\caption{Corrected distributions of the charge-selected EEC compared with the inclusive case (top panel) and charged ratio (bottom panel) for $R_{\rm jet}$ = 0.4 with a jet transverse momentum selection of 20  $ < p_{\rm T, jet} <$ 30 GeV/$c$.  Comparisons with the PYTHIA 8 Detroit Tune \citep{Aguilar:2021sfa} and the default tune of HERWIG 7 are given \citep{Marchesini:1991ch}.}
\label{Tamis_Figure3_0p4}
\end{figure}

Lastly, the charge separated analysis shown in Fig. \ref{Tamis_Figure3} is repeated for $R_{\rm jet}$ = 0.4 in Fig. \ref{Tamis_Figure3_0p4}.  As in the $R_{\rm jet}$ = 0.6 case, an excess of like-sign correlations are seen relative to Monte-Carlo predictions at angles below the transition region, implying hadronization effects not fully captured by current Monte-Carlo models.

\end{document}